# Is the Born rule a result of measurement noise?

Francisco E. Torres

SRI International

frank.torres@sri.com

The Born rule asserts the probability distribution of eigenstates observed in unbiased quantum measurements, but the reason it holds remains elusive. This manuscript discusses how the Born rule might be explained by Schrodinger equation dynamics, if a measurement comprises a system responding to random fluctuations until it is within an arbitrarily small tolerance of a measurement eigenstate. We describe the random walk dynamics that produce this behavior in terms of a class of time-dependent, stochastic unitary matrices U(t). We also discuss the class of stochastic potential energies in Schrodinger's equation that is equivalent to this class of unitary matrices.

This analysis raises some questions worth considering, including how to determine if any measurements actually follow the predicted random walk mechanism and whether a reliable measurement apparatus could be designed that deviates from Born rule probabilities. Interestingly, if any measurements do follow this random walk mechanism, then exposing a quantum system to stochastic "noise" is an intrinsic part of such a measurement, not merely an unwanted side effect. This characteristic would have implications for reducing noise in quantum sensing and quantum computing.

**Acknowledgement:** This work would not have moved forward without encouragement by and discussions with Krishnan Thyagarajan. Comments and questions are welcome.

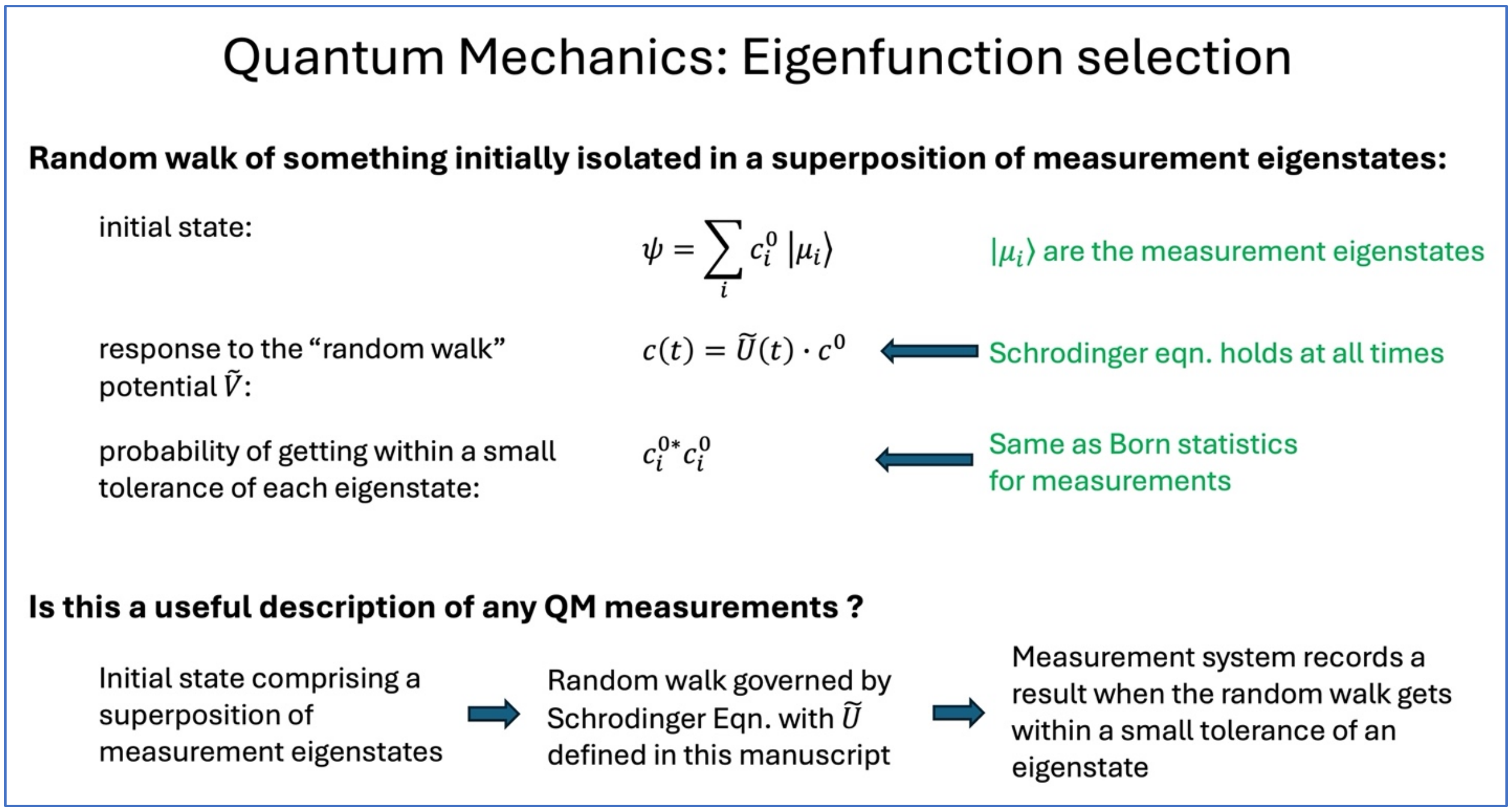

One of the profound results of quantum mechanics is that a measurement necessarily reduces a superposition of measurement eigenstates to one of those eigenstates, following the Born rule for the probability of selecting each eigenstate. In this manuscript, we explore how the Born rule can be explained by Schrodinger equation dynamics, if a measurement comprises a system responding to a certain type of stochastic potential energy until it is within an arbitrarily small tolerance of a measurement eigenstate.

## Definitions

For an observable, consider its measurement operator $\widehat{M}$ having *N* eigenstates $|\mu_i\rangle, i \in \{1,2, \dots, N\}$, normalized such that $\langle \mu_i | \mu_j \rangle = \delta_{ij}$. Consider a quantum system that starts out as a superposition of more than one such eigenstate. Define $\widehat{H}$ to be its Hamiltonian, corresponding to the Schrodinger equation

$$i\hbar \frac{d\psi}{dt} = \widehat{H}\psi \tag{1}$$

Using the eigenstates $|\mu_i\rangle$ as a basis, we can write

$$\psi = \sum_{i=1}^{N} c_i(t) \, |\mu_i\rangle , \;\; where \; \sum_{j=1}^{N} c_j^*(t) c_j(t) = 1. \tag{2}$$

In this manuscript, we restrict attention to measurement eigenstates that are independent of time. Inserting Eq. 2 into Eq 1 and multiplying by $|\mu_k\rangle\langle\mu_k|$ yields

$$i\hbar \frac{dc_k}{dt}|\mu_k\rangle = |\mu_k\rangle \sum_{i=1}^{N} \langle \mu_k|\widehat{H}|\mu_i\rangle \, c_i \, . \tag{3}$$

This manuscript aims to analyze evolution in the measurement eigenstate basis, so define $H$ to be the Hamiltonian matrix in the $|\mu_k\rangle\langle\mu_i|$ basis, i.e. $H_{ki} = \langle \mu_k|\widehat{H}|\mu_i\rangle$, and in the $|\mu_i\rangle$ basis define the column vector

$$c(t) = \begin{pmatrix} c_1(t) \\ c_2(t) \\ \vdots \\ c_N(t) \end{pmatrix}. \tag{4}$$

Then Eq. 3 can be written in the measurement eigenstate basis as

$$i\hbar \frac{dc}{dt} = Hc. \tag{5}$$

For t=0, define

$$c^0 = \begin{pmatrix} c_1(t=0) \\ c_2(t=0) \\ \vdots \\ c_N(t=0) \end{pmatrix} \tag{6}$$

Using matrix exponentiation to solve Eq. 5 subject to the initial condition in Eq. 6,

$$c(t) = e^{-\frac{iHt}{\hbar}} c_0 \tag{7}$$

It is well known that $exp(iA)$ is a unitary matrix whenever *A* is Hermitian, and Hamiltonian matrices are Hermitian. Therefore, Eq. 7 is equivalent to

$$c(t) = U(t) \cdot c^0, \quad U(t=0) = I \tag{8}$$

where $U(t) = exp\left(-\frac{iHt}{\hbar}\right)$ is a unitary matrix in the same $|\mu_k\rangle\langle\mu_i|$ basis as $H$. Eq. 8 is the well know result that a system obeying Schrodinger's equation is restricted to undergoing unitary transformations. Like Eq. 5, Eq. 8 is written in the measurement eigenstate basis, setting us up to analyze state evolution in this basis.

If $U(t)$ is a diagonal matrix for all time, then Eq. 8 implies the Born probabilities $c_i^* c_i = |c_i|^2$ remain the same. Accordingly, we write $U(t)$ as the product of a diagonal unitary matrix *D* and a stochastic unitary matrix $\widetilde{U}$,

$$U = D \cdot \widetilde{U} \tag{9}$$

subject to $D(t=0) = \widetilde{U}(t=0) = I$.

The Hamiltonian matrix *H* in terms of *D* and $\widetilde{U}$ is (see Appendix 3)

$$H = i\hbar \left( D^\dagger \frac{dD}{dt} + \frac{d\widetilde{U}}{dt} \widetilde{U}^\dagger \right), \tag{10}$$

where † denotes the complex transpose. Alternatively,

$$H = H_0 + \tilde{V}, \tag{11a}$$

where we define

$$H_0 = i\hbar\, D^\dagger \frac{dD}{dt} \;\text{ and }\; \tilde{V} = i\hbar\, \frac{d\tilde{U}}{dt} \tilde{U}^\dagger. \tag{11b}$$

$H_0$ is the Hamiltonian in the absence of the stochastic energy $\tilde{V}$, and it alone cannot change the magnitudes $|c_i|$ that determine the relative weights of eigenstates in Eq. 2 because $D$ is a diagonal unitary matrix. The stochastic energy $\tilde{V}$, on the other hand, can cause the relative weights of the eigenstates to change.

*Definition of a wavefunction "reducing to its $k^{th}$ eigenstate":* In this manuscript, a wavefunction "reducing to its $k^{th}$ eigenstate" means it attains a state where the only component in the measurement eigenstate basis is the $|\mu_k\rangle$ component. The coefficient $c_k(t)$ therefore has unit magnitude, meaning it is a complex number $e^{i\theta}$ on the unit circle of the complex plane.

*Defining $\tilde{U}(t)$*: Being a stochastic matrix, the $N \; x \; N$ complex elements of $\tilde{U}(t)$ undergo random movements in the unit circle of the complex plane. However, such movements cannot all be independently random if the matrix is to remain unitary. Indeed, an $N \; x \; N$ unitary matrix has only $N \; x \; N$ degrees of freedom, whereas each element of an $N \; x \; N$ complex matrix has a real and an imaginary part, for a total of $2N^2$ randomly fluctuating components. To continue, we need to specify an appropriate parameterization of $\tilde{U}(t)$.

Referring to Box 1, we parameterize $\tilde{U}(t)$ by specifying the complex numbers $a_k e^{i\phi_{kk}}$ along the diagonal for all but the last row, the phases $\phi_{ij}, i \neq j$, of all off-diagonal elements for all but the last row, and the phase $\phi_{N,N}$ for the diagonal component in the last row. The total number of these parameters is 2(*N*-1)+(*N-1*)(*N-1*)+1=$N^2$, matching the number of degrees of freedom for a $N \; x \; N$ unitary matrix. We further specify the stochastic dynamics to comprise (i) independent, uniformly random trajectories in the unit circle of the complex plane for the *N-1* complex parameters along the diagonal, (ii) independent, uniformly random trajectories in the interval [0,2π) for the off-diagonal phases in rows 1 to *N–1*, and (iii)

$$\tilde{U} = \begin{pmatrix} a_1 e^{i\phi_{11}} & u_{12} e^{i\phi_{12}} & u_{13} e^{i\phi_{13}} & \cdots & \cdots & u_{1N} e^{i\phi_{1N}} \\ u_{21} e^{i\phi_{21}} & a_2 e^{i\phi_{22}} & u_{23} e^{i\phi_{23}} & \cdots & \cdots & u_{2N} e^{i\phi_{2N}} \\ \vdots & \vdots & \vdots & \vdots & \vdots & \vdots \\ u_{N-1,1} e^{i\phi_{N-1,1}} & u_{N-1,2} e^{i\phi_{N-1,2}} & u_{N-1,3} e^{i\phi_{N-1,3}} & \cdots & a_{N-1} e^{i\phi_{N-1,N-1}} & u_{N-1,N} e^{i\phi_{N-1,N}} \\ u_{N,1} e^{i\xi_1} & u_{N,2} e^{i\xi_2} & u_{N,3} e^{i\xi_3} & \cdots & u_{N,N-1} e^{i\xi_{N-1}} & u_{N,N} e^{i\phi_{N,N}} \end{pmatrix}$$

***Box 1****: Unitary matrix parameterization.* The independent parameters are $\phi_{ij} \in [0,2\pi)$ and $a_k \in \mathbb{R}_{\geq 0}$. The values for $u_{ij}$ and $\xi_k$ then follow from the requirement that $\tilde{U}$ is a unitary matrix. For the first row we require $u_{1j} \geq 0$ to eliminate unimportant redundancies. For the other rows, $u_{ij}$ may need to be negative to keep $\phi_{ij}$ and $\xi_k$ in [0, $2\pi$).

an independent, uniformly random trajectory in the interval [0,2π) for the phase of the diagonal component in the last row.

**Lemma 1 – Validity of the parameterization of $\widetilde{\boldsymbol{U}}$**: A unitary *NxN* matrix exists for every set of the independent parameters $\phi_{ij}$ and $a_k$. See Appendix 1 for a proof for *N*=2 and Appendix 2 for a proof for *N*=3. A proof for *N*>3 remains to be done.

## $\widetilde{\boldsymbol{U}}(\boldsymbol{t})$ and evolution of wavefunctions to a single eigenstate

With $\widetilde{U}$ defined, we now discuss how its "random walk" results in the wavefunction getting reduced to a single eigenstate.

If $U(t)$ evolves to a state where its *k^th^* row is the complex transpose of the $c^0$ column vector, then $c(t) = U(t) \cdot c^0$ will have a "1" as its *k^th^* element, since $(c^{0*})^T c^0 = |c^0|^2 = 1$. All other elements of $c(t)$ will be zero, since $c(t)$ retains a norm of 1 under all unitary matrix transformations. The wavefunction $\psi$ in Eq. 2 will then reduce to

$$\psi = |\mu_k\rangle. \qquad (12)$$

This is a special case of the following:

**Lemma 2 - $\boldsymbol{U}(\boldsymbol{t})$ for eigenstates**: The wavefunction $\psi$ in Eq. 2 will reduce to its $k^{th}$ eigenstate at time *t*, i.e.

$$\psi = c_k(t)\,|\mu_k\rangle, \qquad c_i(t) = 0\ \forall i \neq k, \qquad (13)$$

if and only if $c_k(t) = e^{i\theta}$ for some $\theta \in \mathbb{R}$. Furthermore, $c(t)$ satisfies this condition, i.e.

$$c(t) = \begin{pmatrix} 0 \\ \vdots \\ e^{i\theta} \\ \vdots \\ 0 \end{pmatrix} \leftarrow\ k^{th}\ element \qquad (14)$$

if and only if the $k^{th}$ row of $U(t)$ is

$$e^{i\theta}(c_1^{0*} \quad c_2^{0*} \quad \cdots \quad c_N^{0*}). \qquad (15)$$

See the Appendix 1 for a proof of Lemma 2. When $U(t)$ satisfies Eq. 15, we say that it is an eigenstate generator and denote it as $U_k$.

**Corollary**: The diagonal unitary matrix $D$ in $U(t) = D(t) \cdot \widetilde{U}(t)$ does not affect whether the conditions in Lemma 2 are met.

*Proof of Corollary*: Multiplying by any diagonal unitary matrix $D$ does not change the magnitude of the elements of $c(t) = D\widetilde{U}c^0$. It can only change the phase of individual elements, which may impact the value of $\theta$ in Lemma 2 but not whether the conditions of Lemma 2 are met.

Given the above Corollary, we now restrict attention to $U = \widetilde{U}$, the $D = I$ case, and state our measurement hypothesis accordingly.

**Measurement Hypothesis**: In a measurement, $\widetilde{U}(t)$ undergoes its random walk until it satisfies Lemma 2 within a very small tolerance, at which point the $k^{th}$ eigenstate gets "measured". For unbiased measurements, the tolerance needs to be small enough that $\widetilde{U}(t)$ can be expected to broadly sample the space of unitary matrices before coming within tolerance of an eigenstate generator $U_k$.

This hypothesis introduces the notion of a tolerance for the wavefunction being close enough to an eigenstate to trigger completion of a measurement. One can think of this tolerance as a small volume in unitary matrix space surrounding each eigenstate generator. For unbiased measurements, the tolerance volume needs to be the same size for all eigenstate generator points. Otherwise, different tolerance volumes for different eigenstate generator points will result in bias, favoring those eigenstates associated with larger tolerance volumes. We will come back to this point in the Discussion section.

The above "Measurement Hypothesis" also introduces the notion of $\widetilde{U}(t)$ broadly sampling the space of unitary matrices. The random-walk diffusion of $\widetilde{U}(t)$ starts from the $\widetilde{U}(t = 0) = I$ initial condition. If eigenstate generator points nearer to this initial condition have relatively large tolerance volumes, then the random walk has a higher probability of entering these volumes before sampling much of the unitary matrix space, causing bias favoring eigenstate generator points closer to $\widetilde{U} = I$. Conversely, smaller tolerance volumes imply lower probability that a random walk will enter the tolerance volume for an eigenstate generator point nearer to $\widetilde{U} = I$ before broadly sampling $\widetilde{U}$ space. For this reason, the Measurement Hypothesis asserts that the tolerance volume needs to be small enough to allow broad sampling of $\widetilde{U}$ space in a probabilistic sense.

We finish this section by presenting a new Random Walk Postulate:

**Random Walk Postulate**: The probability of the random walk of $\widetilde{U}$ selecting eigenstate *k* for measurement is $\left|c_k^0\right|^2$ in the limit where the Measurement Hypothesis tolerance approaches zero, matching the expected Born rule statistics for measurement of a state initially in a superposition of measurement eigenstates.

The following sections provide drafts of proofs of the Random Walk postulate, first for the simpler *N*=2 case, followed by the more complex $N \geq 3$ case.

## Random Walk postulate for N=2

Referring to Box 1, the independent parameters in $\widetilde{U}$ are $a_1, \phi_{11}, \phi_{12}, and\ \phi_{22}$ for *N*=2. Solving for the rest of $\widetilde{U}$ yields

$$\widetilde{U} = \begin{pmatrix} a_1 e^{i\phi_{11}} & u_{12} e^{i\phi_{12}} \\ -u_{12} e^{i(\phi_{22}+\phi_{11}-\phi_{12})} & a_1 e^{i\phi_{22}} \end{pmatrix}, \tag{16}$$

where the magnitude $u_{12} = \sqrt{1 - a_1{}^2}$. The term $a_1 e^{i\phi_{11}}$ is a complex number in the unit circle of the complex plane, as shown in Fig. 1. The term $u_{12} e^{i\phi_{12}}$ is also shown in Fig. 1.

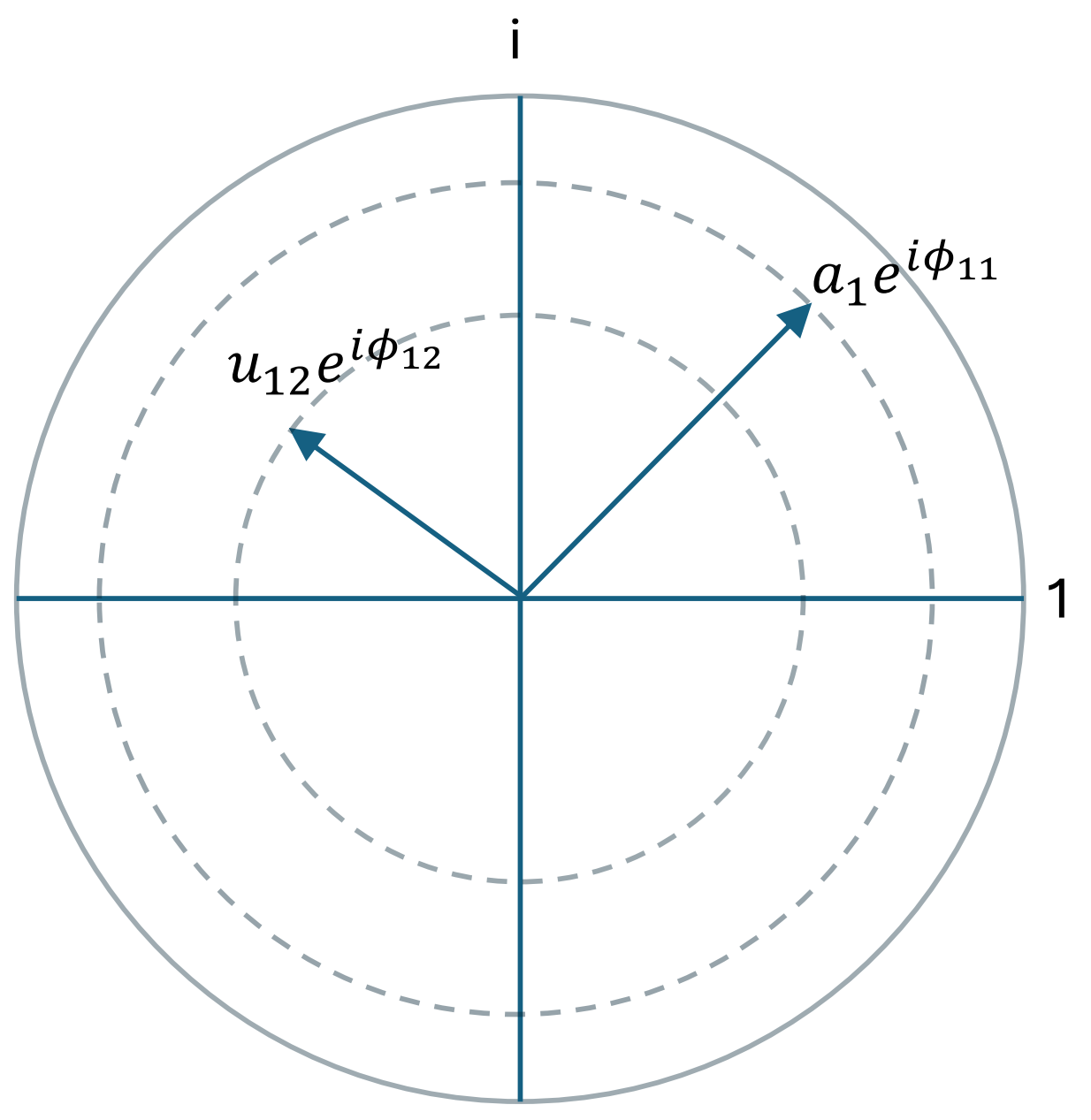

Figure 1: $u_{12} = \sqrt{1 - a_1{}^2}$

Using the Box 1 parameterization raises the question of whether results would be the same if a different parameterization had been chosen. Indeed, different parameter definitions can result in different sampling distributions of $\widetilde{U}$ from the space of unitary matrices, which could very well yield probabilities that differ from those stated in the Random Walk Postulate. For N=2, we note that the Haar measure for our parameterization[1] is

$$a_1\, da_1\, d\phi_{11}\, d\phi_{12}\, d\phi_{22}. \quad (17)$$

The first three terms represent uniform integration over a circle with radial coordinate $a_1$ and angular coordinate $\phi_{11}$ (analogous to $r\, dr\, d\theta$ in physical space), consistent with our sampling $a_1 e^{i\phi_{11}}$ uniformly over the unit circle in the complex plane. The last two differentials are consistent with our sampling of the $\phi_{12}$ and $\phi_{22}$ phases uniformly over the interval [0, 2π). Thus, our specification of $\widetilde{U}$ for *N*=2 complies with Haar measure sampling of unitary matrices, demonstrating that $\widetilde{U}(t)$ draws samples uniformly from the space of 2D unitary matrices, given sufficient time. Any other parameterization that also samples uniformly from the space of 2D unitary matrices should produce the same eigenstate probability results.

Another form of the Haar measure for $U_2$ that is cited often in the literature (e.g. [1]) is

$$d\mu(U) = \sin\theta \cos\theta\, d\theta\, d\phi\, d\alpha\, d\beta \quad (18)$$

[1] Leaving out the normalization constant for simplicity.

for a unitary matrix written as

$$U = e^{i\beta}\begin{pmatrix} \cos\theta\, e^{i\phi} & \sin\theta\, e^{i\alpha} \\ -\sin\theta\, e^{-i\alpha} & \cos\theta\, e^{-i\phi} \end{pmatrix}. \quad (19)$$

Substituting $a_1 = \cos\theta$ and noting $a_1 da_1 = -\sin\theta \cos\theta\, d\theta$ shows that Eqs. 17 and 18 for the Haar measure are equivalent. Ref. [2] is an excellent reference on random matrices, including Haar measures.

Now we analyze the probability $P(\mu_k)$ that $\widetilde{U}$ is in the tolerance volume ε for eigenstate generators defined by Lemma 2, for $k \in \{1,2\}$. For $k$=1, Lemma 2 can be restated as Eq. 13 holding iff the complex numbers in Fig. 1 are a rotation of the pair $(c_1^{0*} \quad c_2^{0*})$. We can factor $P(\mu_1)$ as

$$P(\mu_1) = P_1(c_1^{0*})P_2(c_2^{0*}|c_1^{0*}), \quad (20)$$

where $P_1$ is the probability that $\widetilde{U}_{11}$ is a rotation of $c_1^{0*}$ by some phase angle $\theta$, and $P_2$ is the conditional probability that $\widetilde{U}_{12}$ is a rotation of $c_2^{0*}$ by the same $\theta$ given $\widetilde{U}_{11}$ is a rotation of $c_1^{0*}$ by $\theta$. Referring to Figure 2, a random walk for $a_1 e^{i\phi_{11}}$ that uniformly traverses the unit circle in the complex plane will be at a rotation of $c_1^{0*}$, within a tolerance ε, whenever the random walk is in the ring of radius $|c_1^0|$, width $\varepsilon$, and area $2\pi|c_1^0|\varepsilon$. (Reminder: the magnitudes of a complex number and its conjugate are equal.) The area of the unit circle is $\pi$, so the probability for this to occur is

$$P_1 = 2|c_1^0| \cdot \varepsilon + O(\varepsilon^2). \quad (21)$$

Regarding $P_2$, its precondition $a_1 = |c_1^0|$ implies $u_{12} = \sqrt{1 - |c_1^0|^2} = |c_2^0|$, meaning $\widetilde{U}_{12}$ necessarily lies in the ring of radius $c_2^{0*}$ and width ε shown in Fig. 2 whenever the precondition for $P_2$ is met. When $\widetilde{U}_{12}$ is inside the small circle of diameter $\varepsilon$ in that ring, it is rotated from $c_2^{0*}$ by the same phase difference separating $\widetilde{U}_{11}$ from $c_1^{0*}$. Thus, $P_2$ equals the area of the small circle of diameter $\varepsilon$ divided by the area of the ring at radius $\left|c_2^{0*}\right|$,

$$P_2 = \frac{\varepsilon}{8\left|c_2^0\right|} + O(\varepsilon^2). \tag{22}$$

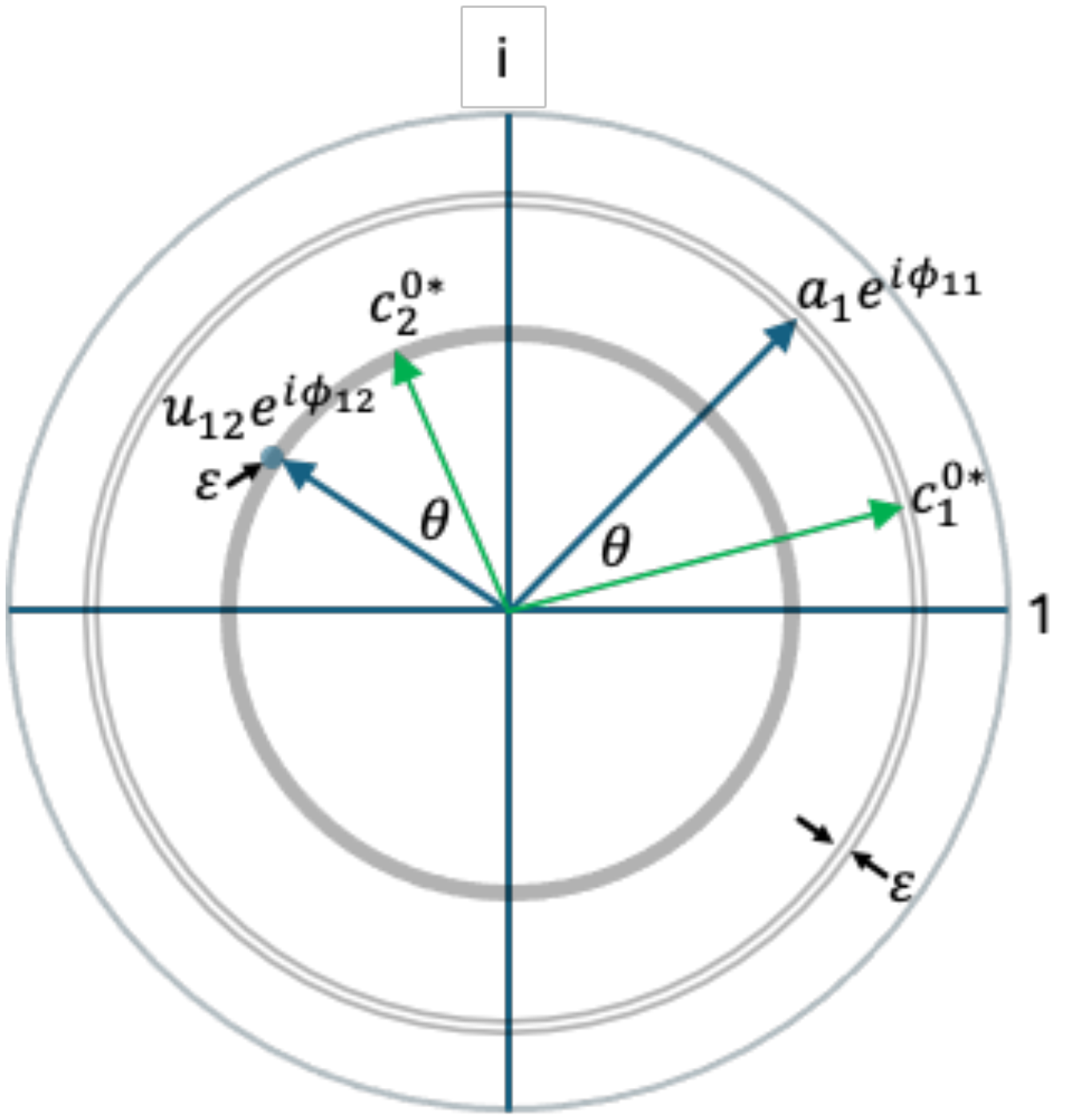

Figure 2: In this example, the first row of $\widetilde{U}$ is a rotation of $(c_1^{0*} \quad c_2^{0*})$ by an angle $\theta$. Note that $\widetilde{U}_{11} = a_1 e^{i\phi_{11}}$ and $\widetilde{U}_{12} = u_{12} e^{i\phi_{12}}$.

Substituting Eqs. 21 and 22 into Eq. 20 reveals

$$P(\mu_1) = \frac{\left|c_1^0\right|\varepsilon^2}{4\left|c_2^0\right|} + O(\varepsilon^3). \tag{23}$$

We can use similar arguments to show that

$$P(\mu_2) = \frac{\left|c_2^0\right|\varepsilon^2}{4\left|c_1^0\right|} + O(\varepsilon^3). \tag{24}$$

The ratio of the probabilities for eigenstate 1 versus eigenstate 2 is

$$\frac{P(\mu_1)}{P(\mu_2)} = \frac{\left|c_1^0\right|^2}{\left|c_2^0\right|^2} + O(\varepsilon) \tag{25}$$

This is the expected Born statistics for an unbiased measurement, equivalent to

$$P(\mu_1) = \left|c_1^0\right|^2 \text{ and } P(\mu_2) = \left|c_2^0\right|^2 \tag{26}$$

because $\left|c_1^0\right|^2 + \left|c_2^0\right|^2 = 1$.

## Random Walk postulate for N>2

For *N*>2, the distributions for $a_i$ in our unitary matrix parameterization are not Haar measure compliant, since they do not have an expected value of 1/*N* except when *N*=2. (See "The Random Matrix Theory of the Classical Compact Groups" by Elizabeth Meckes, ref [2].) Indeed, for *N*>2 the random walk of $\widetilde{U}$(t) is biased towards instances with larger diagonal components.

In this section, we analyze the probability that the random walk of $\widetilde{U}$(t) brings $c(t)$ within tolerance of $e^{i\theta}\psi_k$ for $k \in \{1,2,\dots,N\}$ and some $\theta \in [0,2\pi)$, i.e. determine the probability that the conditions in Lemma 2 are satisfied.

*Definitions and conventions*:

1. $u_1 = \{1^{st}\ row\ in\ \widetilde{U}(t)\} \stackrel{\text{def}}{=} (a_1 e^{i\phi_{11}} \quad u_{12}e^{i\phi_{12}} \quad \cdots \quad u_{1N}e^{i\phi_{1N}})$.
2. All probabilities are understood to mean the probability of being true within some small tolerance $\varepsilon$ in the complex plane.
3. $P(\psi_1) = \{prob\ that\ u_1 = e^{i\theta}(c_1^{0*} \quad c_2^{0*} \quad \cdots \quad c_N^{0*})\ for\ some\ \theta\}$. This is the probability that Eq. 15 in Lemma 2 holds for *k*=1.

We break this probability down into components as follows:

$$P(\psi_1) = P_1 P_{dir} P_2 \prod_{k=2}^{N} P_{\phi_k} \tag{27}$$

$P_1 = \{probability\ that\ the\ diagonal\ component\ a_1 e^{i\phi_{kk}} = c_1^{0*} e^{i\theta}\ for\ some\ \theta, i.e.\ a_1 = |c_1^0|\}.$

$$P_{dir} = \left\{ \begin{matrix} prob\ that\ (a_1, u_{12}, u_{13,} \cdots, u_{1N})\ is\ pointed\ in\ the \\ (|c_1^0|, |c_2^0|, |c_3^0|, \cdots, |c_N^0|)\ direction\ from\ the\ origin\ in\ \mathbb{R}^N,\ \ given\ a_1 = |c_1^0| \end{matrix} \right\}$$

$$P_2 = \left\{ \begin{matrix} probability\ that\ u_{1k} = |c_k^0|\ for\ k = 2, \ldots, N, given\ that\ (a_1, u_{12}, u_{13,} \cdots, u_{1N}) \\ is\ pointed\ in\ the\ (|c_1^0|, |c_2^0|, |c_3^0|, \cdots, |c_N^0|)\ direction\ in\ \mathbb{R}^N\ and\ a_1 = |c_1^0| \end{matrix} \right\}$$

$P_{\phi_k} = \{prob\ that\ u_{1k} e^{i\phi_{1k}} = c_k^{0*} e^{i\theta}\ for\ k = 2, \ldots, N, given\ that\ u_{1k} = |c_k^0|\}$

**Analysis of probabilities**

Starting with $P_1$, recall that $a_1 e^{i\phi_{11}}$ is sampled uniformly from the unit circle in the complex plane for a sufficiently long random walk. There will always be some $\theta$ such that $\theta + (phase\ of\ c_1^{0*}) = \phi_{11}$, so $P_1$ reduces to the probability that $a_1 = |c_1^0|$ within a tolerance of $\varepsilon$. Referring to Figure 2, a random walk that uniformly covers the unit circle in the complex plane will meet the condition for $P_1$ when the random walk is in the ring of area $2\pi|c_1^0| \cdot \varepsilon$. The area of the unit circle is $\pi$. Thus,

$$P_1 = 2|c_1^0| \cdot \varepsilon + O(\varepsilon^2), \tag{28}$$

the same result previously shown for N=2.

Regarding $P_{dir}$, we posit that there is no preference for one direction over another for the vector of magnitudes $(u_{12}, u_{13,} \cdots, u_{1N}) \in \mathbb{R}^{N-1}$ in the prescribed probability distribution for $\tilde{U}(t)$. The statement is trivially true for $N = 2$, since the $(|c_2^0|)$ direction is nothing more than the $\mathbb{R}^+$ direction, i.e. there is only one direction possible. For $N$>2, consider a proof by contradiction as follows: Assume there is a preferred direction over the probability distribution of $\tilde{U}(t)$ , given $a_1 = |c_1^0|$. Now swap two rows j≠1 and $k \notin \{1, j\}$ in all instances of the distribution of $\tilde{U}$ and then swap columns j and k so the diagonal components remain along the diagonal. For at least some choices of j and k, this swapping changes the preferred direction. Nonetheless, the probability distributions of all independent parameters $a_i$ and $\phi_{ik}$ remain the same for all $i = 2, \ldots, N$ before and after the row+column swap, implying the swap of indices does not change the statistical distributions of the components of $\tilde{U}$ and therefore does not change the preferred direction. This is a contradiction, so there must not be a preferred direction for $(u_{12}, u_{13,} \cdots, u_{1N})$.

Since there is no preferred direction for $(u_{12}, u_{13,} \cdots, u_{1N})$ in $\mathbb{R}^{N-1}$, then given $a_1 = |c_1^0|$, the probability that the direction of $(a_1, u_{12}, u_{13,} \cdots, u_{1N})$ in $\mathbb{R}^N$ is within a small tolerance of the direction of $(|c_1^0|, |c_2^0|, |c_3^0|, \cdots, |c_N^0|)$ is the same for all directions of $(|c_1^0|, |c_2^0|, |c_3^0|, \cdots, |c_N^0|)$, i.e. the probability distribution for directions is uniform. It depends only on the tolerance $\varepsilon$. Specifically, the direction of $(|c_1^0|,\ u_{12}, u_{13,} \cdots, u_{1N})$ is within a small tolerance $\varepsilon$ of the direction of $(|c_1^0|, |c_2^0|, |c_3^0|, \cdots, |c_N^0|)$ when these directions intersect the unit sphere in $\mathbb{R}^N$ within a distance $\varepsilon$ of each other. This corresponds to a

finite manifold of dimension N-2 on the unit sphere in $\mathbb{R}^N$ having a size proportional to $\varepsilon^{N-2}$, i.e. loosely speaking,

$$P_{dir} \sim \varepsilon^{N-2} \;\; \text{and therefore} \;\; P_1 P_{dir} \sim \left|c_1^0\right| \cdot \varepsilon^{N-1}. \tag{29}$$

For the interested reader: A more careful analysis of $P_1$ and $P_{dir}$ integrates over all values of $a_1$ within $\varepsilon$ of $\left|c_1^0\right|$:

$$\int_{a_1=\left|c_1^0\right|-\varepsilon/2}^{a_1=\left|c_1^0\right|+\varepsilon/2} \int_{\Omega(\varepsilon)} \hat{P}_{dir}(\Omega; a_1)\hat{P}_1(a_1)\, d\Omega \;\, da_1 \, , \tag{30}$$

where $\hat{P}_{dir}(\Omega; a_1)$ is the probability density for orientation $\Omega$ of $\left( u_{12}, u_{13,} \cdots, u_{1N}\right)$ at a given $a_1$ and $\hat{P}_1(a_1)$ is the probability density for $a_1$. Referring to Fig 2, $\hat{P}_1(a_1) = 2\pi a_1$. Integrating over $\Omega(\varepsilon)$ is defined to mean integrating over all orientations of $\left( u_{12}, u_{13,} \cdots, u_{1N}\right) \in \mathbb{R}^{N-1}$ that satisfy the condition that the direction of $\left(a_1, u_{12}, u_{13,} \cdots, u_{1N}\right) \in \mathbb{R}^N$ intersects the unit sphere in $\mathbb{R}^N$ within a distance $\varepsilon$ of the intersection by the direction of $\left(\left|c_1^0\right|, \left|c_2^0\right|, \left|c_3^0\right|, \cdots, \left|c_N^0\right|\right)$. The locations on the unit sphere in $\mathbb{R}^N$ where the direction of $\left(a_1, u_{12}, u_{13,} \cdots, u_{1N}\right)$ intersects for a fixed $a_1$ and varying $u_{1k}$ (i.e. varying $\Omega$) is an *N*-2 dimensional manifold. The subset of this manifold that is within $\varepsilon$ of $\left(\left|c_1^0\right|, \left|c_2^0\right|, \left|c_3^0\right|, \cdots, \left|c_N^0\right|\right)$ has a size proportional to $\varepsilon^{N-2}$ (ignoring terms that are higher order in $\varepsilon$) when $a_1 = \left|c_1^0\right| + O(\varepsilon)$. Thus, $\hat{P}_{dir}(\Omega; a_1) \sim \varepsilon^{N-2}$, independent of $a_1$ to first order. Consequently,

$$\int_{a_1=\left|c_1^0\right|-\frac{\varepsilon}{2}}^{a_1=\left|c_1^0\right|+\frac{\varepsilon}{2}} \int_{\Omega} \hat{P}_{dir}(\Omega; a_1)\hat{P}_1(a_1)\, d\Omega \;\, da_1 \sim \pi\varepsilon^{N-2}\left[\left(\left|c_1^0\right| + \frac{\varepsilon}{2}\right)^2 - \left(\left|c_1^0\right| - \frac{\varepsilon}{2}\right)^2\right] = 2\pi\left|c_1^0\right|\varepsilon^{N-1}. \tag{31}$$

Summarizing, this more careful analysis returns a result in agreement with Eq. 29:

$$P_1 P_{dir} = \int_{a_1=\left|c_1^0\right|-\frac{\varepsilon}{2}}^{a_1=\left|c_1^0\right|+\frac{\varepsilon}{2}} \int_{\Omega} \hat{P}_{dir}(\Omega; a_1)\hat{P}_1(a_1)\, d\Omega \;\, da_1 \;\sim\; \left|c_1^0\right|\varepsilon^{N-1}. \tag{32}$$

It is relatively easy to see that $P_2 = 1$ to first order: Since $|c^0| = 1$, it follows that $\sum_{k=2}^{N}\left(c_k^0\right)^2 = 1 - (c_1^0)^2$. Likewise, when $a_1 = c_1^0$, $\sum_{k=2}^{N}(u_{1k})^2 = 1 - (c_1^0)^2$ because $u_1$ is a row of a unitary matrix and therefore has a norm of one. Therefore,

$$\sum_{k=2}^{N}(u_{1k})^2 = \sum_{k=2}^{N}\left(c_k^0\right)^2. \tag{33}$$

When the directions of $(u_{12} \quad u_{13} \quad \cdots \quad u_{1N})$ and $(c_2^0 \quad c_3^0 \quad \cdots \quad c_N^0)$ are the same, it follows that all of the ratios $u_{1k}/c_k^0$ are the same, say equal to a constant r, so it is also the case that

$$\sum_{k=2}^{N}(u_{1k})^2 = r^2\sum_{k=2}^{N}\left(c_k^0\right)^2 \tag{34}$$

Eq. 33 and Eq. 34 both being true implies *r*=1 (plus higher order terms in ε), i.e. $u_{1k}/c_k^0 = 1$ for all *k*. Thus, $P_2 = 1$ to first order in $\varepsilon$.

Fig. 2 is a graphical view of $P_{\phi_k}$ for the *N*=2 case, discussed in the previous section. Similarly, for *N*≥2, given the precondition for $P_{\phi_k}$ that $u_{1k} = \left|c_k^0\right|$, the complex matrix component $u_{1k}e^{i\phi_{1k}}$ necessarily lies in the tolerance ring of area $2\pi\left|c_k^0\right|\varepsilon$. For the phase $\phi_{1k}$ to be within ε of $\theta$ in the complex plane requires $u_{1k}e^{i\phi_{1k}}$ to be within an area $\pi\varepsilon^2/4$ subset of this same tolerance ring. Consequently,

$$P_{\phi_k} = \frac{\varepsilon}{8\left|c_k^0\right|} + O(\varepsilon^2), \tag{35}$$

the ratio of $\pi\varepsilon^2/4$ and $2\pi\left|c_k^0\right|\varepsilon$. Note: the constant multiplier is unimportant because we will be taking ratios of probabilities later.

Putting all these probabilities together, including multiplying together all $P_{\phi_k}$ for k=2,…,N, yields

$$P(\psi_1) = constant \cdot \frac{\left|c_1^0\right|^2 \varepsilon^{2N-1}}{\prod_{k=1}^{N}\left|c_k^0\right|} + O(\varepsilon^{N+1}), \tag{36}$$

where we have multiplied the numerator and denominator by $\left|c_1^0\right|$ for later convenience. The same probability analysis can also be done for rows 2 through *N*-1 of $\widetilde{U}(t)$. Therefore, Eq. 36 holds for all $\psi_k, k \in \{1,2,\dots,N\}$.

The result for $P(\psi_N)$, corresponding to the *N*th row of $\widetilde{U}$, requires extra consideration because the magnitude of the diagonal component in that row is not one of the chosen degrees of freedom for the parameterization of $\widetilde{U}$. To investigate the probability distribution for $u_{N,N}$ in the *N*th row, we first investigate the first *N*-1 elements of the *N*th column. Note that the expected value of $a_i^2$ for $i \in \{1,\dots,N-1\}$ is ½. The sum of the expected values of the squared amplitudes for a row necessarily equals one because the norm of each row of a unitary matrix is one, so for each row $i$,

$$\sum_{\substack{k=1\\k\neq i}}^{N} E\left[u_{ki}^2\right] = 1/2\,. \tag{37}$$

Each summand $E\left[u_{ki}^2\right]$ in Eq. 37 should be the same because the prescribed probability distributions of all the off-diagonal parameters for our parameterization of $\widetilde{U}$ are the same. Accordingly, we posit that $E\left[u_{ki}^2\right] = 1/(2(N-1))$. The sum of the expected values of the squared amplitudes in the *N*th row also necessarily equals one because $\widetilde{U}$ is unitary, so

$$1 = E[u_{NN}^2] + \sum_{k=1}^{N-1} E\left[u_{kN}^2\right] = E[u_{NN}^2] - \frac{N-1}{2(N-1)} = E[u_{NN}^2] - \frac{1}{2}. \tag{38}$$

Therefore,

$$E[u_{NN}^2] = \frac{1}{2}\ , \tag{39}$$

which matches the expected value of $u_{NN}^2$ when $u_{N,N}e^{i\phi_{N,N}}$ is uniformly distributed in the unit circle of the complex plane. Thus, like all other rows, we conclude that $u_{N,N}e^{i\phi_{N,N}}$ is indeed uniformly distributed in the unit circle of the complex plane. Regarding the phases $\xi_i$ for $i \in \{1, \dots, N-1\}$, we expect them to be uniformly distributed in [0,2π) because all of the independently prescribed phase parameters for $\widetilde{U}$ are uniformly distributed. Thus, we assert that the probability distribution of the phases in the $N^{\text{th}}$ row is the same as the prescribed distributions for all other rows. In summary, we conclude that the probability distribution for row *N* should be the same as all other rows.

Accordingly, we assert that

$$P(\psi_m) = constant \cdot \frac{|c_m^0|^2\ \varepsilon^{2N-1}}{\prod_{k=1}^{N} c_k^0} + O(\varepsilon^{N+1}) \tag{40}$$

for all $m \in \{1, \dots, N\}$, not just for $m \in \{1, \dots, N-1\}$. The ratio of the probabilities for any two distinct eigenvectors is then

$$\frac{P(\psi_m)}{P(\psi_n)} = \frac{|c_m^0|^2}{\left|c_n^0\right|^2} + O(\varepsilon). \tag{41}$$

This is the expected Born statistics for an unbiased measurement, as $\varepsilon$ approaches zero. Since $\sum_{i=1}^{N}\left|c_i^0\right|^2 = 1$, Eq. 41 is implies

$$P(\psi_i) = |c_m^0|^2, \tag{42}$$

a more familiar version of the Born rule.

## Discussion of implications and applications

We provided a strong argument that a random walk starting from a superposition of eigenstates and obeying Schrodinger's equation at all times can bring a system arbitrarily close to each eigenstate with probabilities matching the expected Born statistics for measurements. But does this actually happen? What experiments could be performed to check?

We leave a proper analysis of the potential $\tilde{V}$ corresponding to our specified random unitary matrix $\tilde{U}$ for a future study. Nonetheless, the question is particularly interesting for *N*>2, since the distribution of $\tilde{U}$ is biased towards the diagonal rather than being Haar measure compliant for all N other than N=2.

If one assumes that a measurement system lets $\tilde{U}(t)$ undergo a random walk while in a superposition of states, and when the system is extremely close to being a single eigenstate the measurement system "holds" or "locks" the system into that eigenstate as part of the measurement (e.g. holding a particle at one position or momentum to get a reading, or converting a photon into an electrical signal to capture an image), then we can predict Born statistics as an outcome of Schrodinger equation dynamics.

If one analyzes the random walk at very short times, one will get different eigenstate probabilities. Because the random walk starts at $\tilde{U}(t=0)=I$, at short times is more likely to find an eigenstate than Born statistics would predict for the larger values of $c_i^0$, and less likely to find the eigenstates corresponding to the smaller $c_i^0$ elements of $c^0$. At such short times, there would also be a significant probability that the random walk would not come within tolerance of any eigenstate. Would this mean there would be a significant probability that no measurement would take place? Are there very short time experiments that could test this theory? The times would have to be very short for a macroscopic measurement system, since quantum times are very small for macroscopic systems. For a macroscopic measurement, this is probably shorter than current experiments could probe. But could an interesting experiment be done wherein the "measurement apparatus" is very small?

In our analysis, we examine getting arbitrarily close to an eigenstate. However, quantum mechanics requires that the system be an eigenstate to get a measurement; arbitrarily close is not good enough. There is still something missing. One possibility is that measurements "hold" or "lock" onto an eigenstate when and only when the quantum system is within some very small tolerance of the eigenstate, and otherwise let the random walk dynamics play out. While this is invoking something outside the Schrodinger equation dynamics discussed here, it is much less drastic than the explanation that something outside of Schrodinger equation dynamics causes the wavefunction to collapse in zero time, undergoing a probabilistic discontinuous change from a superposition of states to a single eigenstate.

If a random walk brings a quantum system arbitrarily close to eigenstates, and then the measurement apparatus takes over by holding onto the eigenstate for long enough to execute the measurement, then it matters whether the measurement apparatus locks into all eigenstates equally. Our analysis assumes this to be true in some sense, in that $\varepsilon$ is assumed to be the same for all eigenstates. But what if some eigenstates have to be approached more closely than others for a measurement to occur? Or what if some eigenstates have a different probability of locking in than others? Extending our analysis to such cases, we would predict deviations from Born statistics. What if we examine a double slit experiment with a detection film that has different probabilities and/or strengths of detection for different eigenstates? Could such an experiment be a test of our theoretical postulate?

In quantum computing, qubits are noisy, and handling that noise is an important issue. If measurements occur by random walks rather than immediate wavefunction collapse, then some “noise”, i.e. the random fluctuations of the random walk, are intrinsic to measuring qubits, not merely a practical annoyance. Can that be tested? If it is true, does it provide new ways to engineer the environment to minimize unwanted effects of noise on quantum computing, i.e. make sure the noise is proper in some sense?

There are a number of questions that our analysis bring up, and it would be interesting to find ways to test some of the questions.

## Appendix 1: Proofs of Lemmas

**Lemma 1**: A unitary *NxN* matrix exists for every set of parameters comprising the complex numbers along the diagonal (both the amplitude and phase) for all but the last row, the phases of all off-diagonal elements for all but the last row, and the phase of the last diagonal term.

*Proof for N=2:* First we show it works for *N*=2. For *N*=2, a unitary matrix can always be written in the form

$$\widetilde{U} = \begin{bmatrix} e^{i\alpha}\cos\theta & e^{i\beta}\sin\theta \\ -e^{-i(\varphi+\beta)}\sin\theta & e^{-i(\varphi+\alpha)}\cos\theta \end{bmatrix}. \quad (A1.1)$$

For the parameterization in Box 1, specifying the diagonal in row 1 sets $\alpha$ and $\theta$ in Eq. A1.1, specifying the phase of the off-diagonal element in row 1 sets $\beta$, and specifying the phase of the diagonal component in the second row sets $(\varphi + \alpha)$. Knowing both $\alpha$ and $(\varphi + \alpha)$ sets $\varphi$. Thus, specifying the independent parameters specified by Box 1 for N=2 sets all parameters in Eq. A1.1. Thus, a unitary matrix $\widetilde{U}$ exists for all choices of independent parameters specified in Box 1 for N=2.

*Proof for N=3:* See Appendix 2

*N*>3: TBD

**Lemma 2 - $\boldsymbol{U(t)}$ for eigenstates**: The wavefunction $\psi$ in Eq. 2 will reduce to its *k*$^{th}$ eigenstate at time *t*, i.e.

$$\psi = c_k(t)|\psi_k\rangle, \;\; c_i(t) = 0 \;\forall i \neq k, \quad \text{(A1.2)}$$

if and only if $c_k(t) = e^{i\theta}$ for some $\theta \in \mathbb{R}$. Furthermore, $c(t)$ satisfies this condition, i.e.

$$c(t) = \begin{pmatrix} 0 \\ \vdots \\ e^{i\theta} \\ \vdots \\ 0 \end{pmatrix} \leftarrow k^{th}\ element. \quad (A1.3)$$

if and only if the *k*$^{th}$ row of $U(t)$ is

$$u_k = e^{i\theta}(c_1^{0*} \quad c_2^{0*} \quad \cdots \quad c_N^{0*}). \quad (A1.4)$$

*Proof:* Assume $c(t)$ is a vector with $e^{i\theta}$ in its *k*$^{th}$ location, $c_k = e^{i\theta}$. Then

$$|c(t)|^2 = \sum_{j=1}^{N} |c_j(t)|^2 = |e^{i\theta}|^2 + \sum_{\substack{j=1 \\ j\neq k}}^{N} |c_j(t)|^2 = 1 + \sum_{\substack{j=1 \\ j\neq k}}^{N} |c_j(t)|^2$$

But $|c(t)|^2 = 1$ at all times, a consequence of $U(t)$ being unitary at all times, so

$$\sum_{\substack{j=1 \\ j\neq k}}^{N} \left|c_j(t)\right|^2 = 0.$$

Since $\left|c_j(t)\right|^2 \geq 0$ for all j and only equals zero when $c_j(t) = 0$, it follows from the above sum being zero that $c_j(t) = 0$ for each $j \neq k$. Thus, if $c(t)$ is a vector with $e^{i\theta}$ in its *k^th^* location, all other elements necessarily equal zero and the wavefunction $\psi$ in Eq. 2 reduces to its *k^th^* eigenstate, i.e. Eq. A1.2 holds. Conversely, assume Eq. A1.2 holds. Since $c_i(t) = 0\ \forall i \neq k$ in Eq. A1.2 and the norm of c(t) is one, $|c_k|^2 = 1$, meaning $c_k = e^{i\theta}$ for some $\theta \in \mathbb{R}$.

Next, assume Eq. A1.4 holds. All rows of a unitary matrix are orthogonal to each other, i.e. $u_j \cdot u_i^* = 0$ for all $i \neq j$, where $u_j$ is defined to be the *j^th^* row of $U(t)$. Thus, when the *k^th^* row of $\widetilde{U}(t)$ is given by Eq. A1.4, then

$$u_j \cdot (c_1^0 \quad c_2^0 \quad \cdots \quad c_N^0)^T = 0 \ \text{ for all } j \neq k \qquad (A1.5)$$

because $e^{i\theta} \neq 0$ for $\theta \in \mathbb{R}$. Eq. 8 states $c_j(t) = u_j \cdot (c_1^0 \quad c_2^0 \quad \cdots \quad c_N^0)^T$, so comparison with Eq. A1.5 establishes that $c_j(t) = 0$ when $j \neq k$. Since the norm of $c(t)$ is always one and all elements of $c(t)$ other than the *k^th^* element are zero, it follows that $c_k(t) = e^{i\theta}$. Thus, Eq. A1.3 necessarily follows from Eq. A1.4. Now we prove Eq. A1.3 implies Eq. A1.4 by proving the contrapositive statement by contradiction. Assume

$$u_k = e^{i\theta}(c_1^{0*} \quad c_2^{0*} \quad \cdots \quad c_N^{0*}) + e^{i\theta} w \qquad (\text{A1.6})$$

for non-zero *w*, i.e. Eq. A1.4 does not hold, yet Eq. A1.3 does hold. Since $u_k$ is a row in a unitary matrix, $u_k \cdot u_k^* = 1$. At the same time,

$$(c_1^{0*} \quad c_2^{0*} \quad \cdots \quad c_N^{0*}) \cdot (c_1^0 \quad c_2^0 \quad \cdots \quad c_N^0)^T = 1$$

from Eq. 2. Therefore, $u_k \cdot u_k^* = 1$ implies

$$(c_1^{0*} \quad c_2^{0*} \quad \cdots \quad c_N^{0*}) w^* + w \cdot (c_1^0 \quad c_2^0 \quad \cdots \quad c_N^0)^T + w \cdot w^* = 0, \qquad (\text{A1.7})$$

When Eq. A1.3 is true, Eq. 8 specifies

$$e^{i\theta} = u_k \cdot (c_1^0 \quad c_2^0 \quad \cdots \quad c_N^0)^T \qquad (A1.8)$$

Substituting Eq. A1.5 into Eq. A1.8 yields

$$e^{i\theta} = \left(e^{i\theta}(c_1^{0*} \quad c_2^{0*} \quad \cdots \quad c_N^{0*}) + e^{i\theta} w\right) \cdot (c_1^0 \quad c_2^0 \quad \cdots \quad c_N^0)^T = e^{i\theta}(1 + w \cdot (c_1^0 \quad c_2^0 \quad \cdots \quad c_N^0)^T).$$

Subtracting $e^{i\theta}$ from both sides and recalling that $e^{i\theta} \neq 0$ for $\theta \in \mathbb{R}$ yields

$$w \cdot (c_1^0 \quad c_2^0 \quad \cdots \quad c_N^0)^T = 0. \qquad (\text{A1.9})$$

Substituting Eq. A1.9 and its complex conjugate into Eq A1.7 yields $w \cdot w^* = 0$, which contradicts the assumption that Eq. A1.6 holds for non-zero *w*. Therefore, if Eq. A1.4 does not hold, then Eq. A1.3 does not hold. The contrapositive must also be true, namely, if Eq. A1.3 holds, then Eq. A1.4 holds. QED.

## Appendix 2: Existence of *U*(*N*=3) for all parameterization values

For *N*=3, the parameters $a_1, a_2, \phi_{11}, \phi_{12}, \phi_{13}, \phi_{21}, \phi_{22}, \phi_{23}$, and $\phi_{33}$ are specified, and we wish to prove that there is a solution for the $u_{ij}$ and $\xi_i$ for all allowed values of the specified parameters.

$$U = \begin{pmatrix} a_1 e^{i\phi_{11}} & u_{12} e^{i\phi_{12}} & u_{13} e^{i\phi_{13}} \\ u_{21} e^{i\phi_{21}} & a_2 e^{i\phi_{22}} & u_{23} e^{i\phi_{23}} \\ u_{31} e^{i\xi_1} & u_{32} e^{i\xi_2} & u_{N,N} e^{i\phi_{N,N}} \end{pmatrix}$$

Focus first on the normalization and orthogonality conditions for rows 1 and 2:

$$a_1^2 + u_{12}^2 + u_{13}^2 = 1 \quad (A2.1)$$

$$u_{21}^2 + a_2^2 + u_{23}^2 = 1 \quad (A2.2)$$

$$a_1 u_{21} e^{i(\phi_{11}-\phi_{21})} + u_{12} a_2 e^{i(\phi_{12}-\phi_{22})} + u_{13} u_{23} e^{i(\phi_{13}-\phi_{23})} = 0 \quad (A2.3)$$

Eqs. A2.1 and A2.2 are the conditions that each row has a norm of 1. Eq. A2.3 specifies that the first and second rows are orthogonal.

Eq. A2.3 maps out a triangle in the complex plane. Since the $\phi_{ij}$ are specified parameters, all possible solutions of Eq. A2.3 are similar triangles. Thus, Eq. A2.3 implies, for a new unknown *p*,

$$\begin{aligned} a_1 u_{21} &= p & (A2.4a) \\ u_{12} a_2 &= s_1(\phi) p & (A2.4b) \\ u_{13} u_{23} &= s_2(\phi) p & (A2.4c) \end{aligned}$$

Different values of *p* correspond to different geometrically similar triangles. The values of the functions $s_1(\phi)$ and $s_2(\phi)$ may be greater than, less than, or equal to zero and are determined by where the phase differences $(\phi_{11} - \phi_{21})$, $(\phi_{12} - \phi_{22})$, and $(\phi_{13} - \phi_{23})$ are pointing in the complex plane.

Eqs. A2.4a-c combined with Eqs. A2.1 and A2.2 provide five equations in the five unknowns $u_{12}, u_{13}, u_{21}, u_{23}$, and *p*. Now we want to show that there is always a solution.

Using Eqs. A2.4a, A2.4b and A2.4c to eliminate $u_{21}$, $u_{12}$ and $u_{23}$ from Eqs. A2.1 and A2.2 yields

$$a_1^2 + \left(\frac{s_1}{a_2}\right)^2 p^2 + u_{13}^2 = 1 \quad (A2.5a)$$

$$\left(\frac{1}{a_1}\right)^2 p^2 + a_2^2 + \left(\frac{s_2}{u_{13}}\right)^2 p^2 = 1 \longrightarrow u_{13}^2 \left(\left(\frac{1}{a_1}\right)^2 p^2 + a_2^2 - 1\right) + s_2^2 p^2 = 0 \quad (A2.5b)$$

Solving for $u_{13}^2$ in Eq. A2.5a, substituting into Eq. A2.5b, multiplying by $a_1^2 a_2^2$, and rearranging as a power series in $p^2$ yields

$$-\left(\frac{s_1}{a_1 a_2}\right)^2 p^4 + \left(\frac{(1-a_1^2)}{a_1^2} + s_1^2 \frac{(1-a_2^2)}{a_2^2} + s_2^2\right) p^2 - (1-a_1^2)(1-a_2^2) = 0. \quad (A2.6)$$

Eq. A2.6 is a quadratic equation in $p^2$, and the discriminant is

$$d = \left(\frac{(1-a_1^2)}{a_1^2} + s_1^2\frac{(1-a_2^2)}{a_2^2} + s_2^2\right)^2 - 4\left(\frac{s_1}{a_1 a_2}\right)^2(1-a_1^2)(1-a_2^2)$$
$$= \left(\frac{(1-a_1^2)}{a_1^2} - s_1^2\frac{(1-a_2^2)}{a_2^2} + s_2^2\right)^2 + 4s_1^2 s_2^2\frac{(1-a_2^2)}{a_2^2}. \qquad (A2.7)$$

Note that $(1-a_2^2) \geq 0$ because $a_2^2 \leq 1$. The squared terms in Eq. A2.7 are positive as well, so $d \geq 0$. Thus, the solutions for $p^2$ are always real.

We also need to show that $p^2 \geq 0$ because $p$ needs to be real. Writing Eq. A2.6 as

$$-Ap^4 + Bp^2 - C = 0, \qquad (A2.8)$$

where

$$A = \left(\frac{s_1}{a_1 a_2}\right)^2$$
$$B = \frac{(1-a_1^2)}{a_1^2} + s_1^2\frac{(1-a_2^2)}{a_2^2} + s_2^2$$
$$C = (1-a_1^2)(1-a_2^2),$$

the roots of Eq. A2.6 are

$$p^2 = \frac{B \pm \sqrt{d}}{2A}. \qquad (A2.9)$$

Note that *A, B* and *C* are $\geq 0$. (Recall that $a_1^2$ and $a_2^2$, the square of magnitudes of complex numbers drawn randomly from within the unit circle of the complex plane, are thereby constrained to be $\leq 1$.) Since $A, C \geq 0$ further implies

$$d = B^2 - 4AC \leq B^2, \qquad (A2.10)$$

both roots given by Eq. A2.9 are greater than or equal to zero.

Next, we need to prove $u_{13}^2$ is greater than zero for all parameterizations, i.e. prove $u_{13} \in \mathbb{R}$. Going back to Eqs. A2.5a and A2.5b to solve for $u_{13}^2$, Eq. A2.5a can be rearranged as

$$p^2 = \left(\frac{a_2}{s_1}\right)^2(1-a_1^2-u_{13}^2) \qquad (A2.11)$$

Substituting this equation for $p^2$ into Eq. A2.5b,

$$u_{13}^2\left(\left(\frac{a_2}{s_1 a_1}\right)^2(1-a_1^2-u_{13}^2) + a_2^2 - 1\right) + \left(\frac{a_2 s_2}{s_1}\right)^2(1-a_1^2-u_{13}^2) = 0. \quad (A2.12)$$

Rewriting as a polynomial in $u_{13}^2$,

$$u_{13}^4 + a_1^2\left(-\frac{1-a_1^2}{a_1^2} + s_1^2\frac{1-a_2^2}{a_2^2} + s_2^2\right)u_{13}^2 - a_1^2 s_2^2(1-a_1^2) = 0. \quad (A2.13)$$

Defining

$$B_u = a_1^2\left(-\frac{1-a_1^2}{a_1^2} + s_1^2\frac{1-a_2^2}{a_2^2} + s_2^2\right),$$

$$C_u = a_1^2 s_2^2 (1-a_1^2),$$

the roots of Eq. A2.13 are

$$u_{13}^2 = \frac{-B_u \pm \sqrt{B_u^2 + 4C_u}}{2}. \qquad (A2.14)$$

Since $C_u \geq 0$, the larger of the roots given by Eq. A2.14 is necessarily positive. Thus, there is always a $u_{13} \in \mathbb{R}$ solution of Eq. A2.12. Given that solution and $p^2$ given by Eq. A2.11, it follows that Eqs. A2.5a and A2.5b are satisfied for this real value of $u_{13}$. Eqs. A2.4a-c then provide a straightforward way to compute $u_{21}$, $u_{12}$, and $u_{13}$.

Finally, after the unknowns in the first two rows of *U* have been determined, the final row is the unique (to within a $\pm 1$ multiplier) remaining orthnormal row with the prescribed phase $\phi_{N,N}$ for the last element.

**Appendix 3: Decomposing the Hamiltonian matrix *H* into a diagonal and a stochastic component**

Eq. 8 introduces the unitary matrix $U(t) = exp\left(-\frac{iHt}{\hbar}\right)$. Its time derivative is

$$\frac{dU}{dt} = -\frac{iH}{\hbar}U. \quad (A3.1)$$

Eq. 9 introduces the decomposition of *U* into a diagonal and a stochastic unitary matrix. Substituting Eq. 9 into Eq. A3.1 yields

$$\frac{dD}{dt}\widetilde{U} + D\frac{d\widetilde{U}}{dt} = -\frac{iH}{\hbar}D\widetilde{U}. \quad (A3.2)$$

Multiplying on the right by $\widetilde{U}^{-1}$ and on the left by $D^{-1}$ yields

$$D^{-1}\frac{dD}{dt} + \frac{d\widetilde{U}}{dt}\widetilde{U}^{-1} = -\frac{iH}{\hbar}. \quad (A3.3)$$

Since $D$ and $\widetilde{U}$ are unitary matrices by definition, $D^{-1} = D^{\dagger}$ and $U^{-1} = U^{\dagger}$, where † denotes the complex transpose. Therefore,

$$H = i\hbar\left(D^{\dagger}\frac{dD}{dt} + \frac{d\widetilde{U}}{dt}\widetilde{U}^{\dagger}\right),$$

which is Eq. 9 in the body of this manuscript.